\documentclass[twoside]{article}
\usepackage[latin1]{inputenc}
\usepackage{t1enc}
\usepackage{a4}
\usepackage{tabularx}
\usepackage{epsf}
\usepackage{graphicx}

\def\bref{\vspace{4pt}\noindent\hangindent=10mm}

\begin{document}

\setcounter{figure}{0} \setcounter{section}{0} \setcounter{equation}{0}

\begin{center}
{\Large\bf
The Early Years: Lyman Spitzer, Jr.\\[0.2cm]
and the Physics of Star Formation}\\[0.7cm]
Bruce G. Elmegreen\\[0.17cm]
IBM Research Division, T.J. Watson Research Center\\
1101 Kitchawan Road, Yorktown Heights, NY 10598, USA\\
bge@us.ibm.com
\end{center}

\vspace{0.5cm}
\begin{abstract}
\noindent{The discovery of the interstellar medium and the early work
of Lyman Spitzer, Jr. are reviewed here in the context of the
remarkable observation in the early 1950's that star formation
continues in the present age. Prior to this observation, stars were
thought to have formed only at the beginning of the universe. The main
debate in the 1930's was whether stars had the young age of $\sim3$ Gyr
suggested by the expansion of the universe and the meteorites, or the
old age of $10^{13}$ yr suggested by thermalized stellar motions. The
adoption of Ambartsumian's claim of modern-day star formation was slow
and mixed in the early 1950's. While some astronomers like Adriaan
Blaauw immediately followed, adding more from their own data, others
were slow to change. By the end of the 1950's, Lyman had deduced the
basic theory for star formation that we would recognize today.}
\end{abstract}

\section{Introduction: Basic Chronology for Star Formation}

In the early 1900's, stars were thought to be ``always'' there, from
the beginning of the universe. No one knew the difference between a
galaxy and a gaseous nebula. Lemaitre's Big Bang theory was not until
1927. In the 1920's, astronomers knew there was an interstellar medium
(ISM, which was discovered in 1904), and in the 1930's, many properties
of this medium began to be measured, but there was no proposed
connection to star formation. This was the context in which Lyman
Spitzer, Jr. received his PhD from Princeton in 1938, working with
Henry Norris Russell on radiative transfer in red supergiant
atmospheres. In the early 1940's, Spitzer developed the first
theoretical models for the ISM, including some ideas on how stars might
have formed long ago; there was still no observation that stars
actually formed today. In 1946, Spitzer wrote his study, ``Astronomical
Advantages of an Extra-Terrestrial Observatory,'' in which he proposed
putting a telescope in space in order to study the uv lines that were
inaccessible from the ground. Finally, in the 1950's, star formation
itself was discovered by Ambartsumian, Blaauw, and a few others. The
real growth in star formation did not begin, however, until the late
1960's and early 1970's, when infrared and CO observations became
possible.

This review will highlight some of the people and discoveries that led
to our modern concept of star formation, with an emphasis on the
contributions by Lyman Spitzer, Jr. A review of the earlier history of
star formation is in Trimble (2009).

\section{Early years of the ISM}

In 1904, Johannes Franz Hartmann (1865-1936) in Potsdam wrote a paper
in the {\it Astrophysical Journal}, ``Investigations on the Spectrum
and Orbit of $\delta$ Orionis,'' in which he observed narrow spectral
lines from this binary system that did not move with the Doppler motion
of the stars. He remarked: ``this point now led me to the quite
surprising result {\it that the calcium line at $\lambda 3934$ does not
share in the periodic displacements of the lines caused by the orbital
motion of the star.}''  The implication was that the calcium line was
not from the stars but from an intervening absorbing cloud. This was
the discovery of the interstellar medium. Edward Emerson Barnard took
photographs of light and dark nebulae at about this same time (Barnard
1908; his famous book of photographs was published posthumously, in
1927), but their gas and dust content was not understood. It would take
20 years before a more complete picture of the extent of the ISM would
be available, following the discovery of pervasive interstellar
reddening from dust.

Among the earliest work associating stars with nebulae was a paper in
1919 by A.J. Cortie, published as a guest lecture for the Royal
Photographic Society entitled ``Photographic Evidence for the Formation
of Stars from Nebulae.'' Cortie noted a morphological sequence of
nebulae arranged by William Herschel in 1811, from faint and diffuse,
to nucleated, then to nebulous stars and single stars in atmospheres of
nebulous light, to groups of stars in nebulous light, and finally to
diffused star clusters and well-defined star clusters. Cortie took
photographic plates of nebulae and commented: ``when the photographic
plate shows thousands of stars involved in nebulous masses, such stars
might be only optically projected upon a background of far removed
nebulous matter, but when the stars are formed on the convolutions of
spirals then the probability becomes overwhelming that the connection
between stars and nebulae is a truly physical one.'' Recall that Cortie
did not know the difference between a gaseous nebula and a galaxy,
which look about the same in a small telescope -- except for the
spirals. He also knew that stars juxtaposed on amorphous nebular
emission could not be certain to have formed there. But he was forced
to conclude, purely from logic, that if stars concentrate in certain
patterns that mimic the patterns of the nebulae, as is the case for
spiral nebulae (i.e., galaxies with their stellar and diffuse light in
spiral patterns), then the stars had to form in these nebula. He took
this as proof that stars are born in nebulae. Of course we know today
that stars and gas concentrate in the spiral arms of galaxies only
partly because of star formation: the arms are usually compression
waves for both gas and stars, which move together in the wave. Why
didn't Cortie conclude that stars associated with diffuse nebulae were
born there too? I think primarily because he did not know the
difference between young and old stars, which required the discovery of
variable and T Tauri stars among the low-mass stars, and short
lifetimes among the high-mass stars. Without knowing this difference,
the numerous stars associated with nebula would all look the same, like
random foreground objects.

An important transition occurred in the early 1920's when Gustaf
Str\"omberg (1925) and others observed differential drift of certain
groups of nearby stars. Str\"omberg's paper: ``The Asymmetry in Stellar
Motions as Determined from Radial Velocities,'' showed that groups of
stars with larger random motions also had larger systematic motions
with respect to the Sun. He did not know the origin of this, but
Lindblad (1927) and Oort (1927) attributed it to galactic rotation,
which was discovered also through the relative motions of globular
clusters and RR Lyrae stars. Str\"omberg and others working on this
drift did not immediately see the connection with star formation --
that came only in the 1950's when a paper involving Spitzer conjectured
that these two stellar populations had different formation histories.
We know now that the second population (Pop I defined by Baade 1944),
with low dispersion and low drift, is the only one actively forming
stars.

The full scale of interstellar matter was not discussed much until Jan
Oort determined an upper limit to the mass column density and average
density near the sun. He reported the results in a 1932 paper entitled
``The Force Exerted by the Stellar System in the Direction
Perpendicular to the Galactic Plane and some Related Problems.'' In
this paper, Oort determined the gravitational acceleration on stars
perpendicular to galactic plane from the distribution of stellar
positions and velocities with height. He compared this acceleration to
the predicted effect from known stars to get an upper limit on the
residual mass that had to be present. He stated: ``We may conclude that
the total mass of nebulous or meteoric matter near the sun is less than
0.05 suns pc$^{-3}$ or $3\times10^{-24}$ g cm$^{-3}$; it is probably
less than the total mass of visible stars, possibly much less.''  This
density corresponds to about 1.5 atoms cm$^{-3}$. In fact, the average
ISM density is about equal to this, perhaps less by 30\%, in modern
surveys. Why did Oort include ``meteoric matter?'' This was something
Lyman Spitzer would also discuss in his ISM course at Princeton
University: meteoric matter throughout space would be invisible in both
emission and absorption and could contribute to the total mass and
gravity without any observable signature. This was long before the
upper limit to the relative baryon content was established from the
theory of Big Bang nucleosynthesis.  The important point from Oort
(1932) was that the gas contributes a small amount to the total disk
mass. To theoreticians at the time, such sparsely distributed gas meant
that gaseous self-gravity could not be very important. As a result,
Spitzer and others were steered away from self-gravity as a trigger for
star formation until the mid 1960's, when disk instabilities were first
investigated by Goldreich, Lynden Bell, and Toomre. It turns out that
disk instabilities can be important, but only on kpc scales.

In 1934, Gustaf Str\"omberg speculated on the origin of galactic
rotation, which was discovered several years earlier in connection with
his observations of differential drift. He wrote a paper in {\it ApJ}
entitled ``The Origin of the Galactic Rotation and of the Connection
between Physical Properties of the Stars and their Motions.'' In this
paper he proposed some foresighted concepts that are still believed
today. First, he suggested that tidal torques spun up pure-gas galaxies
in the early universe, when the galaxies were much closer to each other
than they are now. Viscosity then cooled this gas and made a thin disk.
Note that gaseous cooling was not understood in detail at that time.
The process of collisional excitation and radiation of energy did not
give an ISM temperature until Spitzer's work 6 or 7 years later. Still,
viscosity was known to exist in fluid systems, so interstellar cooling
was discussed in those terms. After disk formation, Str\"omberg
suggested that stars formed. He went on to say that ``calcium vapors
... and dark clouds in the galaxy seem to be the last remnants of the
gaseous material from which the stars have been formed.'' Curiously, he
placed all star formation at the beginning of the galaxy, and none in
today's ``calcium vapors.''

In the same year, Walter Baade and Fritz Zwicky (1934) made a
remarkable discovery that would eventually have implications for nearly
all aspects of interstellar matter, galaxy evolution, and star
formation, but which was virtually ignored and unappreciated for twenty
years. This was the discovery of supernovae. Their first paper, ``On
Super-novae'', was only 5 pages long and was followed in the {\it
Proceedings of the National Academy of Sciences} by another
breakthrough paper ``Cosmic Rays from Super-novae,'' which was also 5
pages long.  Baade and Zwicky noted in their introduction ``The
extensive investigations of extragalactic systems during recent years
have brought to light the remarkable fact that there exist two
well-defined types of new stars or novae which might be distinguished
as `common novae' and `super-novae'.'' They noted that SN1885 in
Andromeda and other galaxies, plus Tycho in 1572 in the Milky Way had
total energies of $\sim10^{51}$ ergs. They then calculated the mass
loss from Einstein's equation $E/c^2$, and concluded ``the phenomenon
of a super-nova represents the transition of an ordinary star into a
body of considerably smaller mass.''  Evidently they were thinking that
the energy loss was from the conversion of mass into energy, whereas in
fact the luminous energy comes from gravitational binding energy of the
collapsed object (plus neutrino energy released from neutron binding
energy during the formation of a neutron star).  They made no
connection (and neither did anyone else for 20 years) between this
energy release and interstellar heating, motion, and compression, and
no one suggested that it triggered star formation during this time
either.  All of these interstellar processes are now believed to be
strongly connected with supernovae. The primary difference then was
that the supernova rate was thought to be much lower than it is today
(by a factor of 10). The breakthrough came with the discovery of
numerous supernova remnants in our Galaxy and other nearby galaxies in
the radio part of the spectrum. Radio astronomy did not begin in ernest
until the early 1950's. Synchrotron emission from the cosmic rays
predicted by Baade and Zwicky was not discovered until Karl Jansky in
1932 and Grote Reber in 1938 first detected it with the earliest radio
telescopes. The emission was not explained until 1947, when Elder et
al. saw an arc of light inside a synchrotron tube in a laboratory at
General Electric. Synchrotron emission had been predicted in 1944 by
Ivanenko \& Pomeranchuk. Today, we have nightly optical supernova
searches in countless galaxies over most of the sky.

Perhaps no discovery so emphatically exposed the interstellar medium as
that of reddening and extinction from dust.  In 1934, Robert Trumpler
of Lick Observatory observed star clusters and found with increasing
cluster distance both an increase in reddening and an increase in
extinction from magnitudes that were too faint for their diameters.
Reddening and extinction can be determined for clusters without the
need for spectra, and because clusters are much brighter than stars,
much larger distances can be covered than with interstellar absorption
lines alone (Trumpler went out to $\sim4$ kpc).  The new results
suggested that interstellar dust was pervasive, with a more-or-less
uniform distribution for as far as it could be seen along the Galactic
plane. Trumpler concluded: ``There seems to be no alternative but to
interpret the observed color excess as being due to selective
absorption of light in space; this will explain not only why the color
excess is always positive but also why we find its largest values in
the most distant clusters.''

It was in this setting, in 1936, that Bart Bok of Harvard College
Observatory sought to establish that all stars formed at the time of
the ``catastrophe'' in the theory of the expanding universe (Lemaitre
1927), which was thought to be $3\times10^9$ yrs ago from the inverse
of Hubble's (1929) expansion constant.  Bok's paper, ``Galactic
Dynamics and the Cosmic Time-Scale'' began by recalling the common
notion that energy equipartition between stars of various types (Seares
1922; Jeans 1934) required a very long timescale, $\sim10^{13}$ years,
for the stellar ages. The equipartition observation was essentially
that stars have a thermal-like velocity distribution function (i.e.,
Gaussian, with low luminosity stars moving faster than high luminosity
stars). Indeed, 2-body relaxation from star-star scattering will
produce such a function, and it will take the incredibly long time
mentioned by Bok. That observation drove the notion that the universe
was old. Bok argued, however, that some star streams are not in
equipartition, and that galactic rotation (Lindblad-Oort) explained why
stars of different types have different properties (i.e., velocity
dispersion versus velocity lag, from Str\"omberg 1925). Bok also said
that star clusters should be dispersing because they are unstable (Bok
1934). (This is true in the sense that the N-body problem has chaotic
orbits for $N\ge3$ and so multiple-star systems must eventually fly
apart. Spitzer was the first to examine the real demise of isolated
clusters, which is evaporation, not instability.) Bok noted that
encounters between globular clusters and clouds should promote the
disruption of those clusters (this pre-dated the Spitzer-Schwarzschild
paper by 18 years). He said furthermore that clusters are unlikely to
form by random coalescence. He did not think they could be very young,
i.e., just formed, but only that they had to be as young as the
Lemaitre expansion age. Finally, Bok said that red giants and late-type
dwarfs coexist in many clusters yet red giants are too luminous to last
more than $10^{10}$ yrs.  Red giants and dwarfs were also known to
coexistent in binaries (Jeans 1929; Kuiper 1935). Thus he concluded:
``The theory of the Expanding Universe indicates that a ``catastrophe''
took place $3\times10^9$ years ago, and it is tempting to place the
origin of stars and stellar systems tentatively at the epoch of this
catastrophe.''  Some of these same arguments would later be used to
suggest that stars formed even more recently than the Big Bang, but
they were never taken as definitive proof of modern star formation
(unlike the observations of expanding OB associations). Curiously, the
age of meteorites was dated to about the same value as the expansion
age (Holmes 1927). So the meteorites, and most likely the Earth, could
be assumed to have formed in the ``catastrophe'' too. Interested
readers should consult a series of articles discussing the age of the
universe from various points of view in {\it Science} Volume 82 (1935).

All of these ideas came before the ISM was systematically studied using
the full variety of atomic transitions. The first of these studies was
by Ted Dunham, who wrote a 1937 paper in {\it PASP} entitled:
``Interstellar Neutral Potassium and Neutral Calcium.'' This paper had
to wait until telescopes were large enough and spectrometers sensitive
enough to see the very faint and narrow lines from interstellar
absorption.  Dunham did his work on a 32-inch Schmidt camera attached
to the coud\'e focus of the 100-inch Mt. Wilson telescope, which was
built in 1917. This was the world's largest telescope from 1917 to
1948. The first real physics was also on the horizon, as Dunham
commented ``An attempt is being made to use the ratio CaI:CaII to
determine the abundance of electrons in interstellar space.'' Ted
Dunham was born in 1897. He got an MD from Cornell in 1925 and a PhD
from Princeton in 1927, eleven years before Spitzer. He also discovered
CO$_2$ in the atmosphere of Venus.

The first glimmer of the existence of giant molecular clouds, which are
known today to be the primary sites of star formation, came in an
article by Jesse Greenstein in 1937, published in the Harvard College
Observatory Tercentenary papers.  The title of his article was ``The
Effect of Absorbing Clouds in the General Absorption Coefficient.'' In
it he writes: ``It is therefore suggested that at least a considerable
part of the observed 'mean absorption coefficient' in low galactic
latitudes arises from absorption by discrete clouds of high localized
absorption.''  That is a pretty slow beginning to GMC research, and
hardly worth noting considering that he did not know how high the
density in these clouds really was, nor how cold the gas was, nor even
that the clouds were primarily molecular. It was part of his PhD
dissertation. Still, others eventually acknowledged the importance of
Greenstein's clouds for on-going star formation, particularly Lyman
Spitzer (1958), as we shall see momentarily.

Jesse Greenstein, born in 1909 in New York City, was one of the most
prominent young astronomers working on interstellar matter in the late
1930's, when Lyman got his PhD. A second was Bengt Str\"omgren, born in
1908 and the son of another Danish astronomer, Elis Str\"omgren. Bengt
Bengt Str\"omgren grew up during the heyday of the fantastic new theory
of quantum mechanics with a reputation as a brilliant student and Niels
Bohr as a family friend. At age 30, Bengt wrote a paper for the {\it
Astrophysical Journal}, ``The Physical State of Interstellar
Hydrogen,'' in which he discussed the recent discovery by Struve and
Elvey (1938) of Balmer emission lines from ionized nebulae. He states:
``It is found that the Balmer-like emission should be limited to
certain rather sharply bounded regions in space surrounding O-type
stars or clusters of O-type stars.''  This conclusion contrasted with
the notion put forward by one of the most eminent astronomers of the
day, Sir Arthur Eddington, who said the ISM should not be significantly
ionized because Hydrogen is so strongly absorbing that the radiation
cannot escape the vicinity of a star. Str\"omgren reasoned however,
that the highly ionized region close to the star would be transparent
to the Lyman continuum stellar radiation, which could therefore
penetrate much further into space than the optical path length for
neutral absorption. Only after a long path of partially neutral
hydrogen would enough total absorption occur to remove completely the
ionizing light, and at that point, the transition to neutral would be
very sudden. Thus he concluded that space would be divided into fully
ionized gas and virtually non-ionized gas, with little volume in
intermediate form.  We all learn about Str\"omgren early in our studies
because of the ionized regions that bear his name.

\begin{figure}
\centering
\includegraphics[width=3in]{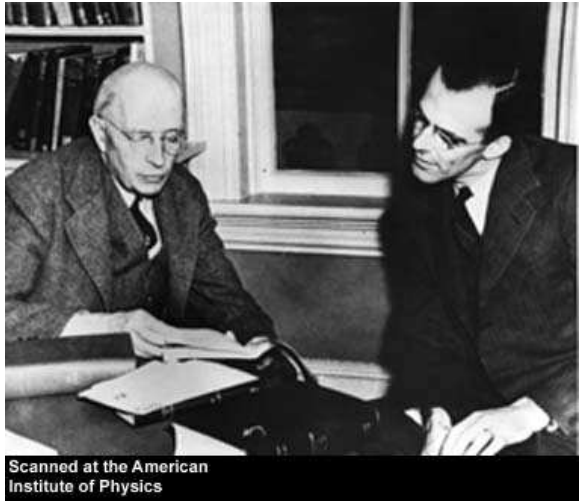}
\caption{Lyman Spitzer, Jr. (right) and Henry Norris Russell, his
thesis advisor, most likely in Princeton in the early
1940's.}\label{fig:1}\end{figure}

\section{Lyman Spitzer's Early Work}

At this point, Lyman Spitzer, Jr., enters the scene. Lyman was born in
1914 and raised in Ohio as the son of a prominent box manufacturer.
There is still a Spitzer Building in downtown Toledo, a Registered
National Landmark, that comes from the family business. Lyman attended
the private school Andover and then Yale as an undergraduate, after
which he went to Cambridge University for a year to study with Sir
Arthur Eddington. There he wrote a paper on ``Non-Coherent Dispersion
and the Formation of Fraunhofer Lines'' (1936). He returned to the
United States to get his PhD in 1938 at Princeton University, working
with Henry Norris Russell, one of the most respected theoreticians of
the time (see Figure 1). Lyman's thesis was on ``The Spectra of Late
Supergiant Stars'' (1938a). This was followed by similar papers ``New
Solutions of the Equation of Radiative Transfer'' (1938b) and ``Spectra
of M Supergiant Stars'' (1939a). These are interesting papers because
his analysis suggested the existence of a slow wind from red
supergiants. Thinking about these stars would also have gotten him into
the same frame of mind as Bart Bok three years earlier, i.e., that the
stars are so luminous they must be short lived. Yet Lyman does not
comment on this as evidence for on-going star formation until later.

Lyman Spitzer's first paper on something like the ISM was also among
the first to apply gas kinetic theory to a topic in star formation. In
1939(b) he published ``The Dissipation of Planetary Filaments'' in
rebuttal of the theory by Lyttleton (1936) and Jeffreys (1929) in which
the planets were proposed to have formed in tidal filaments drawn out
from the Sun during a near-collision with another star. This planet
formation theory was not as crazy as it might seem today. Recall that
the stars were thought to be $10^{13}$ years old from the observation
of kinetic equilibrium (Bok's paper refuting this was not until 1936)
and yet the Earthly rocks and meteorites were known to be only several
Gyr old from radioactive decay. Thus the planets had to form long after
the sun, and there was plenty of time during this long wait, $10^{13}$
years, for a near collision with another star to occur. Such collisions
would indeed make tidal debris, and this debris could in principle,
cool to make the planets. Henry Norris Russell (1934) suggested the
debris was from a former binary companion that was disrupted during an
encounter with a third star. What Lyman showed was that for either case
the debris coming off a star was so hot, and the density so low, that
the gas would expand into the vacuum of space before atomic collisions
and subsequent radiation could cool it. It would not form planets, but
would only evaporate. This rebuttal apparently had considerable
influence because the theory seemed to have dropped from general
discussion, except for counter rebuttals by Lyttleton over the next few
years.

In 1940, at age 26 and only 2 years from his PhD, Lyman made one of his
most important discoveries and started a new field of research that
would lead to many papers and the classic book {\it Dynamical Evolution
of Globular Clusters} (1988). This was his paper ``The Stability of
Isolated Clusters.'' In this paper he notes that energy equipartition
means a star cluster should establish an isothermal velocity
distribution, in which case it should extend to infinity if it has no
boundary pressure. ``How can this be?'' he asks. The answer is that the
cluster evaporates over time. Random motions populate the high velocity
tail of the isothermal Gaussian distribution function, and the stars in
the tail, beyond some cutoff, have a speed that exceeds the escape
speed from the cluster.  He used the $\pi/2$ deflection time from Jeans
(1929) and Smart (1938), along with the virial theorem, to calculate
the rate of mass loss. He also considered centrally condensed clusters
and a range of stellar masses. The timescale he derived for a typical
galactic cluster would be rather short if only the visible stars were
used, but he did not consider this case. Instead he showed that the
timescale would be long ($10^{12}$ yrs), as required to save
appearances in the popular model, if the average star mass were small,
0.1~M$_\odot$. For example, he noted that NGC 2129 would have a
relaxation time of $\sim10^7$ yrs, and therefore had to form less than
the commonly accepted value of $10^9$ yrs ago, unless it lost half of
its mass and had few low-mass stars left.  Today, we consider the short
lives of clusters to be one of the many proofs that star formation is
continuous, and we use cluster age distributions to derive star
formation histories. How could Lyman have missed what is so obvious to
us today? He simply did not know the stellar initial mass function,
which was not discovered until 1955.

The history of our understanding of star formation is an example, like
many others in science, of an incredible resistance to new ideas during
a transition time when old ideas, however absurd they appear to us now,
could not be clearly disproved, and new ideas, however obvious they are
to us now, could not be unambiguously demonstrated.  There were many
signs in the 1940's and even earlier that stars had to form
continuously, but they were not even seen, let alone recognized, by
most astronomers before the discoveries of important empirical laws
(e.g., stellar mass functions) and certain key astronomical objects
(e.g., radio Supernova Remnants, Giant Molecular Clouds, pre-main
sequence stars, pre-stellar cores, etc.). Knowing what we know now, we
can see clues about star formation in even the earliest observations
(Herschel's nebulae, for example), but either no one could see these
clues at the time or no one felt bold enough to speculate in print
about a world view so different than the norm. In the 1940's, they were
still wondering how the Universe could be as young as the Hubble
expansion time and yet have equipartition in the motions of stars.

Spitzer's next paper introduced another whole field of research that
would soon have a direct bearing on star formation, although not in a
way that led to its ready acceptance as a current process. This was his
paper on ``The Dynamics of the Interstellar Medium: I Local
Equilibrium'', published by the {\it Astrophysical Journal} as the
first of three parts on this general topic in 1941-1942. Lyman was at
Yale University when he wrote this trilogy. In this first paper, he
derived the negative charge on dust, the ISM viscosity, and the
timescale for dust drag through gas. He also introduced the terms HI
and HII regions. He made the interesting comment: ``although
equipartition of energy probably exists in any small region of
interstellar space, large-scale turbulence or galactic currents are not
ruled out for this reason,'' which indicates he was thinking about
turbulence, but not very enthusiastically.

In fact it was the same year that the Russian statistician A.N.
Kolmogorov published ``The Local Structure of Turbulence in
Incompressible Viscous Fluid for Very Large Reynolds Numbers.'' This
was the beginning of the famous ``Kolmogorov law'' of relatively
velocity scaling as the 1/3 power of relative distance.  Also in this
year, Fritz Zwicky published ``Reynolds Number for Extragalactic
Nebulae,'' which, like Spitzer's paper, contained a derivation of the
viscosity and a comment on turbulence. However, Zwicky was a bit
louder, saying in the abstract, ``turbulent flows will play a major
role in the morphology of nebulae.'' Here he means ``nebulae'' as
galaxies because he says ``The first indispensable cornerstone for the
hydrodynamic analysis of nebulae is laid through a discussion of the
range of Reynolds' number $R$ both for stellar systems proper and for
the systems of interstellar gases which may be present in the form of a
matrix of a stellar system.''  The consideration of turbulence as an
integral part of the story of the ISM, in terms of power sources, gas
motions, cloud structures, and even the formation of clouds and stars,
always seemed far from Lyman's mind.  This will arise on other
occasions too, as we shall see momentarily.

In the second of his 1941 trilogy, entitled ``The Dynamics of the
Interstellar Medium: II Radiation Pressure,'' Lyman considered the
effect of radiation pressure from field stars on dust. He showed that
dust and gas are usually coupled, and that the radiative force on dust
is stronger than the radiative force on gas. Then he proposed that dust
particles are pushed together by background starlight through a
shadowing effect in which each particle in a pair shadows the other,
giving an inverse square law of mutual attraction. In fact, the
radiative force attracting two dust particles was calculated to exceed
200 times the gravitational force between them. Combining this with
what was then thought to be the gas-to-dust mass ratio of 10 (instead
of 100 as we believe today), he concluded that radiation pressure
dominates gravitational self-attraction in the ISM.  This was his
mechanism of cloud and star formation -- radiative attraction between
dust grains -- to which he adhered for more than ten years. Lyman
claimed: ``Such a force is clearly of great importance in the formation
and equilibrium of condensations within the medium.''

Before dismissing this idea too quickly, we should consider again the
observations of the time. Interstellar gas was viewed as extremely
rarefied, so self-gravity in the gas alone was weak, except on galactic
scales, and then rotation and shear interfere. This is still true for
the formation of the largest clouds ($10^7\;M_\odot$), which contain
GMCs as subunits. The difference today is that we take cloud
fragmentation for granted (Hoyle's theory on fragmentation was proposed
12 years after 1941), and we understand how azimuthal forces,
especially with an azimuthal magnetic field, can mitigate the influence
of rotation and shear (from the 1980's and 1990's). But to Lyman in
1941, the ISM seemed too low in density to have made stars at the
beginning of the universe when it had to do that, and any assistance
from radiation pressure was welcome.

The third of Lyman's trilogy, published in 1942, was entitled ``The
Dynamics of the Interstellar Medium: III Galactic Distribution.'' This
was a remarkable paper. He determined the heat dissipation rate in the
ISM from atomic collisional processes and showed that cooling is so
rapid the gas cannot maintain a 3D structure in elliptical galaxies.
Therefore the diffuse light in ellipticals had to come from numerous
unresolved stars, not from scattering off of dust grains. He then
wondered what happened if the cool gas settled to the center of the
elliptical galaxy. Using the virial theorem, he derived a maximum gas
mass for stability in the presence of a distribution of stars. This
derivation is like that for a pressurized isothermal sphere (which was
not derived until 15 years later), but with background stellar gravity
replacing the cloud boundary pressure as a containing force. If the
nuclear gas cloud is too massive, it cannot support itself and must
collapse. He also derived in this paper the more-famous ${\rm
sech}^2(z/z_0)$ solution for the vertical equilibrium of density in a
self-gravitating isothermal sheet. He proposed that if the cooling gas
in an elliptical fell to a plane, then the scale height of the ISM in
that plane would be 10-100 pc. ``Such a medium would probably be
unstable and would perhaps condense into stars, meteorites, or large
dark bodies,'' he concluded. He then went on to propose that a thin
layer of ``dark matter'' exists in elliptical galaxies, and he derived
the resulting vertical profile for starlight above this layer,
comparing it with observations (from Oort 1940).

\begin{figure}
\centering
\includegraphics[width=3in]{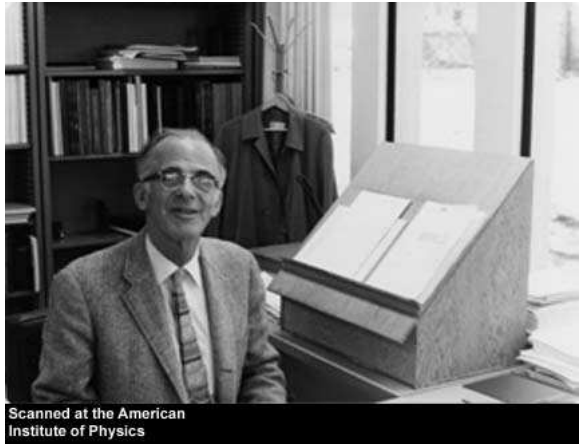}
\caption{Lyman Spitzer, Jr. in his office at Princeton University, when
I was his student, around 1973. This was the time of the {\it
Copernicus} satellite, which observed interstellar absorption lines in
the ultraviolet.}\label{fig:2}
\end{figure}

Over the next few years, Lyman returned to the topics of radiative
transfer in stellar atmospheres, the temperature of interstellar
matter, and charged dust. These were the war years when he worked
mostly on the theory of sonar. In 1946, following the development of
the V2 rocket in Germany, he published his famous paper for the RAND
corporation ``Astronomical Advantages of an Extra-Terrestrial
Observatory.''  In it he cites the importance of ground state
absorption transitions in interstellar atoms, which are mostly in the
uv and can be seen only from space, and the desirability of getting
above the turbulent atmosphere to avoid blurred images. This eventually
led to the Orbiting Astronomical Observatory series of telescopes, of
which Lyman had the third, named {\it Copernicus}. It was launched in
1972 (1 year from the 500th anniversary of the birth of Copernicus). I
was a graduate student then at Princeton, and {\it Copernicus} caused
quite a stir in the department when it found widespread molecular
Hydrogen in diffuse clouds, OVI absorption from $10^5$~K gas, elemental
depletion in dust, and many of the ground-state transitions from
important atoms that Lyman had predicted nearly 30 years earlier. His
second prediction, clear images from space, would not be realized for
another 21 years, when the Hubble Space Telescope received its first
servicing mission (in 1993, corrective mirrors were inserted to
compensate for figuring errors in the primary). Project Stratoscope
(1957-1971) came first, taking clear images from the top of the
atmosphere with a 12'' mirror initially and then a 36'' mirror hanging
from a Balloon. Lyman's colleague at Princeton, Martin Schwarzschild,
ran the project (Fig. 3).

\begin{figure}
\centering
\includegraphics[width=3in]{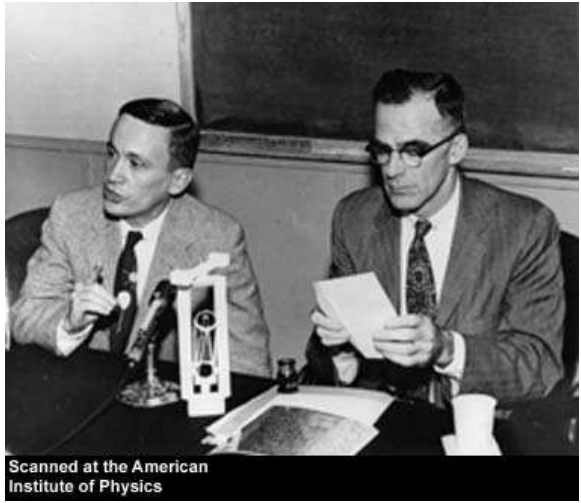}
\caption{Martin Schwarzschild (left) with Lyman Spitzer, Jr. at a
demonstration of Project Stratoscope, probably in the late 1950's. A
model of the telescope gondola is on the table. Note the Stratoscope
and Sun image on Martin's necktie.}\label{fig:3}
\end{figure}

There was one important paper during the war years that should be
mentioned in this chronology, that of Walter Baade in 1944 entitled:
``The Resolution of Messier 32, NGC 205, and the Central Region of the
Andromeda Nebula.''  Here, Baade described photographs of these objects
taken for the first time with red-sensitive plates, showing resolution
into stars. The resulting Hertzsprung-Russell diagram for these stars
was found to be the same as that for globular clusters in the Milky
Way.  He stated: ``This leads to the further conclusion that the
stellar populations of the galaxies fall into two distinct groups, one
represented by the well-known H-R diagram of the stars in our solar
neighborhood (the slow-moving stars), the other by that of the globular
clusters. Characteristic of the first group (type I) are highly
luminous O- and B-type stars and open clusters; of the second (type
II), short-period Cepheids and globular clusters. Early-type nebulae
(E-Sa) seem to have populations of the pure type II. Both types seem to
coexist in the intermediate and late-type nebulae.'' Baade noted that
Jan Oort had recognized these two populations in the Milky Way in 1926.
The importance of this observation could not be lost on Lyman Spitzer
and others who had long recognized that Type I stars always occur in
conjunction with gaseous nebulae and extinction, as in the spiral arms
of galaxies, while Type II stars never do.

Thus, in 1948, and again in 1949, Spitzer wrote reviews on ``The
Formation of Stars.''  First repeating a lot of what Bart Bok said in
1936, Lyman stated that the prevailing ideas needed revision. All stars
were supposed to begin at the same mass and burn down over time, but
this could not be right. He claimed: the universe is too young for most
stars to radiate away their matter; Uranium on the Earth cannot be
explained by contamination from the sun, and some clusters should not
last long before evaporation. This all meant ``something happened about
three billion years ago. If the universe was not created then, it was
certainly very extensively reorganized; some sort of cosmic explosion
apparently took place at that time.''  He went on to note the high
luminosity of supergiant stars, stating ``We conclude that these
supergiant stars have probably formed within the last hundred million
years.''

Then Spitzer presented other evidence for star formation: supergiant
stars found only near interstellar clouds; supergiant stars found only
in spiral galaxies, and clouds of matter between the stars found only
in spiral galaxies, especially Andromeda. He stated, ``the
observational evidence indicating a physical connection between clouds
and supergiant stars is very strong.'' And then came ``a plausible
picture for the birth of a star: ... start with an interstellar gas,
formed at the same time as the rest of the universe... the first step
in the process is then the slow condensation of interstellar grains ...
after these grains have reached a certain size, the radiative
attraction between them forces them together and they drift toward each
other, forming an obscuring cloud in a time of about ten million years
... the radiative force becomes ineffective when the clouds become
opaque ... at this point, gravitation takes over ... small opaque
clouds of this type, called globules, have been known for some
time...''

Clearly he was still thinking in the context of the prevailing notion
that most star formation began at the time the galaxy formed. Thus
grains had to form first. He made a singular exception for supergiant
stars, which he said formed from gas clouds within the last several
hundred million years.  He was also still using his radiation pressure
theory, which we know today is incorrect. But the basic steps of cloud
formation followed by star formation are present in his writing. He
also noted ``One of the chief problems concerns the angular momentum of
this prestellar globule ... perhaps turbulent motions carry the angular
momentum away... '' (he attributed this to ``German astronomers,''
probably meaning von Weizs\"acker as described below). ``In this
country the possibility has been advanced that a galactic magnetic
field might produce electrical eddy currents in a rotating protostar,
which would then damp out the angular momentum ...'' (this probably
came from the observation of interstellar polarization, published by
Hiltner and Hall in 1949). Lyman later worked with Leon Mestel of
Sussex University (1956) on angular momentum damping by magnetic fields
in protostars.

Fred Whipple added to Lyman's ideas in an important way. In 1946,
before Lyman's review articles but at a time when his work must have
been well known outside the published literature, Whipple wrote a paper
for {\it ApJ} entitled ``Concentrations of the Interstellar Medium.''
In it he said that ambient radiation pressure is far too slow as a
formation mechanism for stars in the general ISM. He proposed instead
the same model for the interiors of Taurus-like clouds, where the
extinction is about 1 magnitude. The mechanism was still slow, as it
required extremely slow protostellar velocities (meters per second) for
condensations to sit around long enough to collect the gas in this way,
but it was much better than in the ambient medium. Thus Whipple
concluded that ``stars formed by concentration of the interstellar
material would be expected to occur in clusters rather than singly.''
He was thinking of supergiant stars for the case of current star
formation, like Spitzer. He also noted that the Taurus cloud masses and
velocities are similar to those of clusters. This was the first
suggestion that young stars prefer the cluster environment and that
known interstellar clouds have the right general properties to make the
known star clusters. Yet we have to ask why this radiation pressure
theory lasted so long if it required such extreme conditions. The
answer apparently goes back to the lack of a basic calibration of the
dust-to-gas mass ratio. The gas mass could not be measured from the
available optical absorption lines without a lot of assumptions about
radiative and collisional equilibria. In fact, the gas-to-dust ratio
was first derived after radio astronomy began, using the HI line at
21-cm to get the gas mass. Arthur Edward Lilley (1955) did this for his
PhD Dissertation at Harvard College Observatory. Lilley's publication
acknowledged Spitzer for ``helpful discussions.''

We should also ask why Lyman considered the recent formation of only
supergiant stars. This again stemmed from the lack of any concept of an
initial mass function, not to mention the even greater mystery that we
cannot solve today, why the IMF is so constant in the local universe.
Without the IMF (and without knowledge of pre-main sequence tracks on
the HR diagram, pre-main sequence winds, pre-main sequence variability,
and everything else about star formation that we have learned in the
mean time), only the most obvious super-luminous stars could be claimed
to be young in the 1940's.

The year 1946 also contained another remarkable paper on star formation
by Kenneth Essex Edgeworth, an engineer by profession and, curiously,
independent discoverer of the Trans-Neptunian (Kuiper Belt) objects.
Edgeworth lived in Ireland and wrote in MNRAS ``Some Aspects of Stellar
Evolution -- I.''  Here he outlined what is essentially today's theory
of star formation, jumping decades over Lyman Spitzer's work, but then
promptly dismissed it for lack of any corroborating evidence. ``The
theory of gravitational instability fixes a minimum distance between
condensations, and it is shown that this rule involves the conclusion
that the stars must have been formed at a temperature very little above
the absolute zero, a result which appears to be highly improbable.'' He
also noted that the angular momentum from galactic rotation is too
large to form the solar system and stated in his second paper that star
formation by successive condensations (i.e., hierarchical
fragmentation) overcomes these problems. However, in that case stars
should form in enormous clusters ($10^5$ stars), which they don't (he
said). Finally, in paper III he suggested that the rotating gas disk of
a galaxy, at a temperature of $\sim10^3$~K, breaks up into azimuthal
filaments, which break up into stars following the removal of heat. The
residual material around each star makes planets.  Let's examine these
ideas in turn. First, it is commonly acknowledged today that stars form
by predominantly self-gravitational forces in the ISM, starting with
cloud formation on kpc scales, and then with self-gravity contributing
to the formation of GMCs, cluster-forming cores in GMCs, and even
individual stars. This gets all the way down to individual stars
precisely because the temperature where stars form is nearly absolute
zero, perhaps 8K in a typical pre-stellar core. Second, hierarchical
fragmentation did not become popular until Hoyle (1953), but Edgeworth
noted that it solved the temperature problem: with higher densities
closer to the final act of star formation, extremely low temperatures
are not so important to get the characteristic unstable mass in the
range of a single star. However, hierarchical fragmentation means that
stars should form in hierarchical patterns (which we know today is
true) and also that stars should form in giant complexes of
$10^4\;M_\odot$ or more (which we also know now to be true). Edgeworth
ruled out both scenarios for lack of evidence. Finally, his preferred
model overcame angular momentum by first forming azimuthal filaments,
which then fragment along their lengths into stars. Much of star
formation is in fact filamentary (e.g., Teixeira, Lada \& Alves 2005),
and the giant galactic filaments he envisioned are a lot like spiral
arms. So Edgeworth was essentially right because he did the appropriate
theory correctly, but he could not have known he was right because
there were no observations of the key predictions of his model. He
would have to wait 24 years for the discovery of ultracold CO in Orion.
He died two years after this CO discovery, by the way, in 1972. (For a
biography of Edgeworth, see Hollis 1996).

The influence of Bart Bok on the formative phase of modern star
formation should not be underestimated. While he and others were
debating the ages of stars and the mechanisms of primordial star
formation, Bart was also investigating local clouds that appeared most
likely to be sites of actual star formation -- then or now. In his
paper with Edith Reilly in 1947, they said ``In recent years several
authors have drawn attention to the possibility of the formation of
stars from condensations in the interstellar medium (Spitzer 1941b,
Whipple 1946). It is therefore necessary to survey the evidence for the
presence in our galaxy of relatively small dark nebulae, since these
probably represent the evolutionary stage just preceding the formation
of a star.''  These words are not followed by any clarification as to
whether they meant the current formation of low-mass stars (as opposed
to supergiant stars which Spitzer and Whipple discussed), or the
primeval formation of low-mass stars. In any case, Bok eventually
proposed that low-mass stars form today in such clouds, and he studied
them for many years, finding confirmation of his ideas in the late
1970's (e.g., Bok 1978). Bart Bok died at age 77 in 1983.

\section{A Prolific Scientist in the 1950's}

Spitzer's work to establish the general properties of the diffuse ISM
reached a pinnacle in the 1950s, at which point he began to divert his
attention to plasma physics and the Stellarator (Eliezer \& Eliezer
2001) at first, and then space telescopes later on. In 1950, Lyman
wrote ``The Temperature of Interstellar Matter: III,'' in which he
summarized the important temperatures and densities: HII regions have a
temperature of $\sim10,000$~K, HI regions, 60~K to 200~K, the average
atomic density is $\sim1$ cm$^{-3}$, the density inside diffuse clouds
is $\sim10$ cm$^{-3}$, and the density between the clouds is $\sim0.1$
cm$^{-3}$. These and other observations of the diffuse ISM were the
basis of his famous book {\it Diffuse Matter in Space} (1968). He also
referred to Greenstein (1946) for the density of the Orion nebula:
$10^2-10^3$ cm$^{-3}$. An interesting comment in Spitzer's paper was
that ``a substantial abundance of H$_2$ must be considered when the
dust density exceeds $10^{-12}$ cm$^{-3}$.'' He would later be the
first to prove this with his {\it Copernicus} observations. (Carruthers
[1970] discovered interstellar H$_2$ with rocket-born telescopes, but
detailed measurements by Spitzer's team in the mid-1970's using {\it
Copernicus} showed clearly the conditions under which H$_2$ would
form).

Also in 1950, Lyman wrote a famous paper with Walter Baade on ``Stellar
Populations and Collisions of Galaxies,'' in which he proposed that S0
galaxies are stripped of their gas by direct collisions between former
spirals. He argued that the collision speed would be so large in a
cluster of galaxies that the individual stellar orbits in each galaxy
would hardly deflect; only the gas would be affected by ram-pressure
shocking. Of course, we know now that direct galaxy collisions make a
big mess with the stellar orbits changing so completely that an
elliptical galaxy can result. (Stars scatter in the changing potentials
of the whole galaxies, not just off other stars as Spitzer proposed.)
This paper also stated, however, that additional stripping could come
from the collision between a cluster spiral and the gas debris from
former galaxy collisions. This is in fact what is likely to happen, if
we consider that the debris he mentions is seen today as part of the
hot intracluster medium (e.g., Kenney, van Gorkom, \& Vollmer 2004).

Another interesting paper, ``A Theory of Interstellar Polarization,''
was published in 1951 with John W. Tukey, a fellow Princeton professor
who claimed greater distinction later as the co-inventor of the Fast
Fourier Transform (with James Cooley in 1965). This paper exclaimed
that ``the polarization of starlight from distant stars, found by
Hiltner and Hall, is perhaps one of the most unexpected discoveries of
modern astrophysics.'' The authors proposed that polarization is caused
by the alignment of iron particles in grains, much like iron filings
align parallel to the field lines surrounding a bar magnet. A
contemporary theory by Davis and Greenstein (1951) is now better
accepted: grains damp their energy while conserving angular momentum in
the local field direction. The result is an alignment of elongated
grains perpendicular to the field, around which they spin like
propellers.

A spate of papers in 1951 and 1952 contained several key discoveries: a
study of CH and CH$^+$ molecules in diffuse clouds (Bates \& Spitzer
1951), 2-photon emission from Hydrogen in planetary nebulae (Spitzer \&
Greenstein 1951), stellar scattering by giant cloud complexes (Spitzer
\& Schwarzschild 1951), metallicity differences between high and low
velocity stars (Schwarz\-schild, Spitzer, \& Wildt 1951), observations
of interstellar sodium (Spitzer \& Oke 1952), discovery of the
variation of the ratio of calcium to sodium with cloud velocity (Routly
\& Spitzer 1952), and supersonic motions in diffuse clouds (Spitzer \&
Skumanich 1952). Also in 1952, Lyman published ``The Equations of
Motion for an Ideal Plasma,'' (Spitzer 1952), which began his long-time
development of the Stellarator Fusion Device at the Forrestal Campus in
Princeton.

Another important paper in 1951 should be mentioned at this point.
Baron C.F. von Weizs\"acker (1912-2007), one of the independent
inventors of the CNO fusion cycle in stars, published a paper ``The
Evolution of Galaxies and Stars,'' in which he outlined an ISM model
where turbulence plays a central role.  He applied it to the time of
galaxy formation, so like others, he was not proposing that stars form
today. Nevertheless, it was foresighted in its scope. He said, for
example, ``Gas in cosmic space is moving according to hydrodynamics,
mostly in a turbulent and compressible manner.'' He included the
Kolmogorov law of velocities, discussed the resulting hierarchy of
clouds, the formation of flattened, spinning, centrally condensed
disks, and the competition between cosmic expansion and turbulence
compression. He also said ``Irregular nebulae must be young, spirals
intermediate, elliptical nebulae generally old. Spiral structure is the
distortion of turbulent clouds by nonuniform rotation.''  These
statements would pretty much fit in with modern discussions. However,
he stuck to the prevailing notion by also saying that ``young stars,''
by which he meant those in spiral arms, ``seem to be, more exactly,
rejuvenated stars,'' which means they accrete interstellar matter to
replace their fusion loss. They are not really young: ``Stars could be
formed as long as there were no stars present, because stellar
radiation inhibits the contraction of clouds to form new stars.''
Still, he explained correctly that stars rotate because of cloud
turbulence, they lose their angular momentum by ``magnetic-hydrodynamic
process,'' and they form planetary or double-star disks in the process.
He identified the characteristic mass for stars as the thermal Jeans
mass for temperatures that are several degrees above absolute zero.

Von Weizs\"acker also proposed that giant stars have condensed cores
with slowly expanding atmospheres, and that they form planetary
nebulae. He said ``giants can be old stars. It is therefore necessary
to find a model for at least some giants that will provide an energy
source which will outlast nuclear energy sources. The only sufficient
source known in the physics of today is gravitation. Enough
gravitational energy can be available only if the giant contains a
highly condensed core... The radiation pressure must drive the
atmosphere outward.... A planetary nebula would then be a special type
of giant in which the atmosphere happens to be transparent to visible
light. As far as I know, no calculations for such models have as yet
been made.'' This is essentially today's model of planetary nebulae.

The turbulent model of the ISM gained some following in papers by von
Hoerner (1951), Chandrasekhar \& M\"unch (1952), Minkowski (1955), and
Wilson et al. (1959), but Spitzer did not follow their lead. In his
1954 rocket-effect paper with Oort entitled ``Acceleration of
Interstellar Clouds by O-type Stars,'' the authors investigated the
origin of cloud motions. They wrote against the turbulent model: ``Von
Weizs\"acker (1948, 1949) has proposed that the motions can be regarded
as turbulence generated by differential galactic rotation, and he
suggests that the spread of relative velocities with a region will vary
as the cube root of the size of the region, as in isotropic turbulence.
... However, there is some question as to whether the theory of
isotropic turbulence in an incompressible fluid can actually be applied
to supersonic motions of a gas with large density fluctuations in a
rotating galaxy. ... The picture we obtain looks rather different from
what would be expected on the theory of turbulent motions. Instead of
more or less contiguous vortices, we find concentrated clouds that are
often separated by much larger spaces of negligible density. Whenever
two such clouds collide, there will be considerable losses of kinetic
energy, and it is uncertain whether the transfer of energy from the
rotation of the Galaxy .. to smaller and smaller eddies.. can maintain
the velocities of the clouds at their observed level.''  Oort and
Spitzer were correct in their reasoning: the ISM is not a swirl of
eddies, and gas collisions contain shocks that dissipate energy all at
once, not in a steady cascade to smaller and smaller scales.  However
ISM turbulence is strongly driven (by supernovae, among other things),
and the compressed regions (shells, clouds) hit each other and fracture
multiple times, making a hierarchical structure anyway (see models in
de Avillez, \& Breitschwerdt 2007). These processes would have been
impossible to visualize and quantify before modern computer
simulations. Moreover, the enormous energy they require would not be
evident for another 20 years, when OVI was discovered by Spitzer's {\it
Copernicus} telescope (Jenkins \& Meloy 1974). Only then was the
enormous influence of supernovae on the dynamics of the interstellar
medium fully appreciated (e.g., McKee \& Ostriker 1977).

An important puzzle regarding diffuse cloud motions, discovered by
Routly \& Spitzer in 1951, was finally solved by Spitzer in his 1954
paper ``Behavior of Matter in Space.'' There he explains why the ratio
of sodium to calcium in diffuse cloud absorption lines varies with
cloud speed: ``in the low velocity clouds calcium atoms are mostly
locked up in the grains, while the sodium atoms are not. ... The normal
ratio in high velocity clouds could then be attributed to general
evaporation or dissociation of the grains produced at the same time
that the clouds were accelerated.'' Many of the results produced by the
{\it Copernicus} satellite in the mid-1970's would involve measurements
of gas depletion onto grains (see review by Spitzer \& Jenkins 1975).

\section{The Beginnings of Modern Star Formation}

We have followed the history of Lyman Spitzer's studies of ISM and star
formation, which was mostly primordial star formation, up to the point
where the key discovery of modern-day star formation was made. Viktor
A. Ambartsumian began publishing his ideas in the Russian journals in
1948 and 1949, and presented them to the Western world at the Colloque
International d'Astro\-physique tenu à Li\'ege in 1953. Here he stated:
``only the detailed study of O- and T- associations made it clear that
a simultaneous origin of stars in groups is a general rule... as long
as stellar associations are systems with positive total energy, i.e.,
unstable and expanding systems, their lifetime must be rather short (of
the order of $10^7$ years for O-associations and $10^6$ years for
T-associations).'' Then he calculated the ratio of the galaxy age to
the association age, and showed from this that all stars in the galaxy
could have formed in associations {\it continuously} over time.  This
final step was the key piece of theory that tied together the whole
concept of modern-day star formation. There was no ambiguity anymore.

Viktor Ambartsumian was born in 1908 in Tbilisi, which is now the
capital of the Republic of Georgia. In 1932 he published ``On the
Radiative Equilibrium of a Planetary Nebula'' in the {\it Bulletin de
l'Observatoire central Poulkovo}, laying the groundwork for the theory
of gaseous nebulae. In 1936 he solved a problem in the conversion of
stellar radial speeds to physical speeds, which led to his State Prize
of the Russian Federation in 1995. During the Second World War, he
solved problems of radiative transfer in media with variable
scattering, absorption, and index of refraction using novel invariance
principles.  He was awarded the Stalin Prize for this in 1946. For his
discovery of expanding OB associations, we was awarded another Stalin
Prize in 1950. Ambartsumian lived until 1996.

Ambartsumian's results on expanding OB associations were known to
Adriaan Blaauw before the Li\'ege conference in 1953, as Blaauw wrote
about them in his 1952 paper entitled ``The Age and Evolution of the
$\zeta$ Persei Group of O- and B-type Stars.'' There Blaauw stated:
``The hypothesis of the occurrence of such expansive motions has been
recently introduced by Ambartsumian in order to explain the presence of
the large but sparsely populated groups of early-type stars, called
`O-associations' by this author (Ambartsumian 1949).'' Blaauw derived
an expansion age for the Per OB association of 1.3 Myr, which is small
compared to the nuclear age of these massive stars, $< 7$ Myr. Blaauw
also noted the Barnard dark clouds nearby and claimed ``a genetic
relation with the group is very probable.'' Thus we have by 1952 the
concept of current star formation in dense dark clouds.

Earlier indirect indications of current star formation can easily be
found. In 1948, for example, George Herbig wrote his PhD Dissertation
on ``A Study of Variable Stars in Nebulosity.'' He was studying the T
Tauri class of stars, as discovered by Joy (1945), and other variable
stars in nebulae. However, Herbig did not mention that these stars
could be young until his 1952 paper in {\it JRASC}, in which he said
``If these are normal stars, the question arises if they are ordinary
field objects that have moved into the nebulae and exhibit emission
spectra and light variations as a result of some interaction process
with the nebular material, or whether they are young stars that have
been formed within the clouds. The Russian astronomers V. Ambartsumian
and P. Kholopov, in their studies of stellar associations, have given
the name ``T-associations'' to the groups of irregular variables in the
dark clouds of Taurus, Corona Australis, and elsewhere. They believe
that the T-associations are composed of newly formed, main-sequence
stars, which in the early stages of their lives behave as unstable
objects with irregular variations in light. Some new evidence that may
support this hypothesis of Ambartsumian and Kholopov will be presented
later, but it will be seen that this evidence is somewhat ambiguous.
The question may be regarded as the most fundamental one in the subject
at the present time.''  The ambiguity, Herbig later claimed, derived
from the possibility that the peculiar emission from these stars comes
from their interaction with the surrounding nebulosity, rather than the
stellar photospheres. There are also many other stars in typical
T-Tauri regions that do not have peculiar spectra.

The transition to full acceptance of the star formation hypothesis was
rather mixed around the world.  By 1953 \"Opik proposed immediately
that stellar associations got their expansion from supernova-induced
star formation, and in 1955, Oort proposed the expansion came from
HII-region induced triggering. The right answer about expansion came
from Zwicky (1953), however: ``As the principle cause for the eventual
expansion of the cluster we propose phenomena which are capable of
dispersing the gas cloud. As a consequence of this dispersal the little
star cluster must of necessity expand itself because the velocity
dispersion which it acquired (Virial theorem) in its stationary
association with the gas cloud cannot be held in check by the much
reduced mass of the star cluster alone.'' He went on to suggest that
supernovae disrupt the cloud. In fact, pre-main sequence winds and HII
regions mostly disrupt star-forming clouds, but stellar dispersal
follows from this cloud disruption more than anything else. Disruption
also follows from a second process that Zwicky named: ``Uneven
dispersal of the enveloping gas cloud by one of the agents mentioned
will be quite sufficient to accelerate the imbedded stars outward to
terminal velocities as high as those observed.'' That is, cloud
disruption leads to moving cloud pieces, and embedded stars follow
these pieces into the surrounding association. However, Zwicky used his
theory of expanding associations to assert that the stars did not have
to form in the cloud, that they could have been old stars that accreted
cloud mass to become early-type stars.

Still, there was no stopping the recognition of OB associations as an
important class of objects. In 1953, Morgan, Whitford and Code
discovered the spiral arms of the Milky Way using the positions of OB
associations, and Guido M\"unch showed through their velocities that
gas clouds also follow these spirals. Blaauw, Herbig, and many others
quickly adopted the star formation model and found ready explanation
for peculiar spectra and variability, clustering associated with
nebulosity, and other previous observations.

The theory of stellar evolution was converging on the idea of young
stars at the same time.  Edwin Salpeter (1952) discovered the triple
alpha fusion process in red giant cores, and noted the plausible
formation of heavier elements inside massive stars, following Hoyle
(1946). In offering an explanation for the newly discovered
(Schwarzschild, Spitzer, \& Wildt 1951) difference in metallicity for
Population I and II stars, Schwarzschild \& Spitzer (1953) correctly
stated ``We may tentatively conclude that white dwarfs, just as the
heavy elements in Population I stars, are the products of nuclear
reactions occurring during the life of the Galaxy.'' But they were
still thinking that this enrichment, like star formation itself,
happened long ago: ``It would not appear to us too surprising if heavy
stars had been much more common in the early days of the Galaxy...It
seems possible that the bulk of the heavy elements, as well as the
white dwarfs, were caused by the deaths of massive stars, mainly in the
early phase of the life of our Galaxy.'' Even so, they finally gave up
on the claim that radiation pressure was the cause of star formation,
noting that the metallicity effect was too large to be explained by
differential dust drift. Yet, Greenstein in 1956 commented on the same
observation: ``We have evidence that the interiors, as well as the
surfaces of the stars of Population II, are poor in C,N,O and the
metals.'' He referenced Chamberlain \& Aller (1951), Martin and Barbara
Schwarzschild (1950), and Nancy Roman (1950). Greenstein repeated the
standard view: ``it seems likely that the universe is older than the
solar system, but the difference may be only half a billion years.''
And then goes on, ``perhaps the metals concentrated to the galactic
plane before the stars were formed, so that Population II stars were
formed from relatively hydrogen rich material. Or perhaps Population I
stars are formed from interstellar matter containing a larger fraction
of dust grains relative to gas...'' One year later, Burbidge, Burbidge,
Fowler \& Hoyle (1957) wrote their ground breaking review on the
``Synthesis of the Elements in Stars,'' in which they laid out in
detail the modern theory of nucleosynthesis and element formation.

This time in the early 1950's was also the beginning of radio astronomy
with the 21-cm line of Hydrogen, discovered by Ewen \& Purcell (1951)
and Muller \& Oort (1951).  A remarkable discovery connected with HI
was made by Bark Bok, who wrote in 1955: ``Radio Observations (21-cm)
of Dense Dark Nebulae.'' Bok found ``inside a dark complex the
variations in density of interstellar dust are not accompanied by
parallel variations in the density of neutral hydrogen... There remains
the possibility that the neutral hydrogen in the dark centers is mostly
in molecular form.'' In this same year, Lilley determined the
gas-to-dust mass ratio by comparing the HI column densities in diffuse
clouds to the extinctions.

In the next year, 1956, Spitzer had two important papers, one with
Mestel (1956a), mentioned earlier, on the removal of angular momentum
from protostars by magnetic fields, and another (1956b) on the need for
hot gas at high galactic latitude, which he suggested followed from the
observation of neutral clouds there and the requirement that these
clouds have some boundary pressure.  He also wrote two years later
(1958a) ``Distribution of Galactic Clusters,'' which was another in his
series of papers on internal cluster dynamics.

\begin{figure}
\centering
\includegraphics[width=3in]{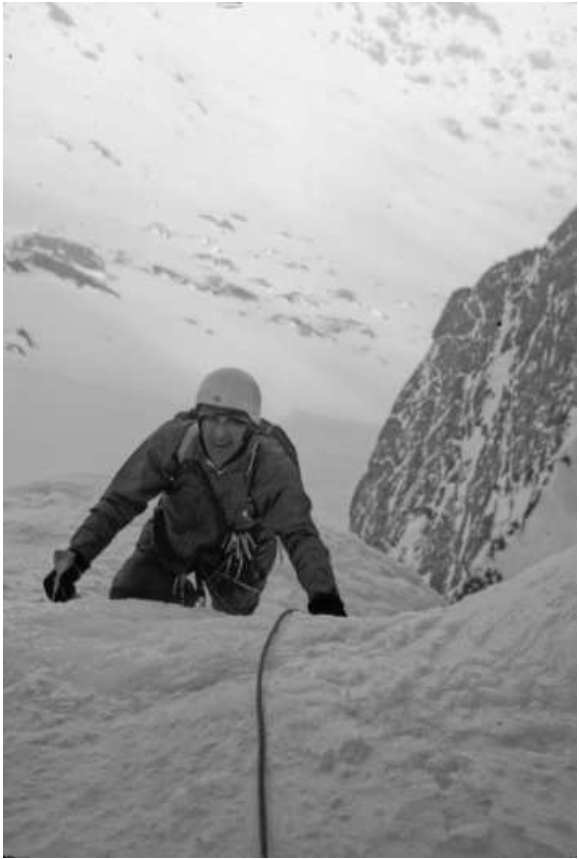}
\caption{Lyman on a difficult climb up Mt. Washington (USA),
photographed by Donald Morton, his colleague at Princeton University.
Lyman and Donald were avid mountaineers, who would entertain graduate
students at dinner parties with slide shows of their amazing
adventures.}\label{fig:4}\end{figure}

In spite of all the ups and downs on the theory of star formation
during this important transition period, Lyman got it pretty much
correct by the end of the decade following Ambartsumian's great
discovery. In 1958, Lyman (1958b) wrote ``Mass Exchange with the
Interstellar Medium and the Formation of Type I Stars,'' for the
Vatican Observatory. He considered the known cloud types and stated
``the so-called 'large cloud' is similar to .. the structures listed by
Greenstein.'' Here he referred to the high absorption clouds we
mentioned above, from Greenstein (1937). He continued, ``it is entirely
possible that in a large cloud, moving more slowly, the internal
velocity dispersion may be less than 1 km/sec.'' He then considered the
mass-to-radius ratio in these clouds and the virial theorem to conclude
``subregions of a large cloud, with low internal velocity dispersion,
might start to contract. Moreover, as the turbulence decays,
gravitational contraction will presumably begin... Altogether it is
entirely reasonable to assume that a typical large cloud ultimately
becomes the birthplace of many galactic clusters and associations.''
Thus he showed that the Greenstein clouds could be virialized, at which
point gravitational contraction to stars could begin.

Nine years later, Becklin \& Neugebauer (1967) wrote ``Observations of
an Infrared Star in the Orion Nebula,'' in which they used a new
infrared detector at Mount Wilson observatory. They concluded the
long-awaited discovery by stating ``An attractive interpretation of the
observation is that the infrared object is a protostar.''  Three years
later, Wilson, Jefferts \& Penzias (1970) discovered ``Carbon Monoxide
in the Orion Nebula.'' These two papers opened up the modern field of
star formation. A new generation of astronomers was suddenly able to
observe star formation in action, creating a revolution in the way we
view and interpret interstellar processes. This revolution continues
today through the construction of an enormous interferometer for
mm-wave emission from CO and other molecules related to star formation
(the ALMA project in Chile), and through the Spitzer Space Telescope,
which observes in the infrared. Lyman Spitzer did not join this CO and
IR revolution. He continued to publish papers on the diffuse ISM for 29
more years.

There was a tale in the halls of Princeton University that there were
two Lyman Spitzers. In the 1950's and 1960's, his work on plasma fusion
at Forrestal Lab off the main campus carried on nearly full time in
parallel with his work on space astronomy and interstellar physics at
Peyton Hall. Somehow he seemed to be both places at the same time. The
students, though, were mostly oblivious to these two great projects,
and far more impressed by his adventures in mountain climbing (Fig.
\ref{fig:4}). Indeed, there were probably three Lyman Spitzers -- maybe
more. But this is the beginning of the legend and the end of the
history. We can save the legend for another time.

Lyman Spitzer collected 39 of his papers with commentary in the book
{\it Dreams, stars, and electrons: selected writings of Lyman Spitzer,
Jr.}, published by Princeton University Press in 1997 (co-authored with
Jeremiah Ost\-riker of Princeton). A recent history of the Hubble Space
Telescope, including many tales about Lyman and his colleagues, is in
{\it The Universe in a Mirror: The Saga of the Hubble Space Telescope
and the Visionaries Who Built It}, by Robert Zimmerman (Princeton
University Press, 2008).

\section{Bibliography}

{\small

\bref Ambartsumian, V.A. 1933, Publication: Bulletin de l'Observatoire
central Poulkovo, 13

\bref Ambartsumian, 1949, Soviet AJ, 26, 3

\bref Ambartsumian, V. A. 1954, Les Processus Nucl\'eaires dans les
Astres, Communications pr\'esent\'ees au cinqui\`eme Colloque
International d'Astrophysique tenu \`a Li\`ege, p. 293 (this conference
took place in September 1953).

\bref Baade, W. 1944, ApJ, 100, 137

\bref Baade, W., \& Zwicky, F. 1934, Proc. Nat. Acad. Sci., 20, 254

\bref Barnard, E.E. 1908, AN, 177, 231

\bref Bates, D.R., \& Spitzer, L.Jr. 1951, ApJ, 113, 441

\bref Becklin, E. E., \& Neugebauer, G. 1967, ApJ, 147, 799

\bref Blaauw, A. 1952, BAN, 11, 405

\bref Bok, B.J. 1934, Harvard College Circular, 384, 1

\bref Bok, B.J. 1936, Obs, 59, 76

\bref Bok, B.J. 1978, PASP, 90, 489

\bref Bok, B.J., \& Reilly, E.F. 1947, ApJ, 105, 255

\bref Burbidge, E. M., Burbidge, G. R., Fowler, W.A., \& Hoyle, F.
1957, RvMP, 29, 547

\bref Carruthers, G.R. 1970, ApJ, 161, 81

\bref Chamberlain, J.W., \& Aller, L.H. 1951, ApJ, 114, 52

\bref Chandrasekhar S., \& M\"unch, G. 1952, ApJ, 115, 103

\bref Cooley, J.W., \& Tukey, J.W. 1965, Math. Comput. 19, 297

\bref Cortie, A.L. 1919, Obs, 42, 398

\bref Davis, L., Jr., \& Greenstein, J.L. 1951, ApJ, 114, 206

\bref de Avillez, M.A., \& Breitschwerdt, D., 2007, ApJ, 665, L35

\bref Dunham, T. 1937, PASP, 49, 26

\bref Elder, F.R., Gurewitsch, A.M., Langmuir, R.V., \& Pollock, H.C.
1947, Phys.Rev., 71, 829

\bref Eliezer, S., \& Eliezer, Y. 2001, The Fourth State of Matter: An
Introduction to Plasma Science, CRC Press, p. 167

\bref Ewen, H. I., \& Purcell, E. M. 1951, Nature, 168, 356

\bref Greenstein, J. 1937, Annals of the Astron. Obs. of Harvard Col.
vol. 105, no. 17, p. 359

\bref Greenstein, J.L. 1946, ApJ, 104, 414

\bref Greenstein, J.L. 1956, PASP, 68, 185

\bref Hollis, A. J. 1996, J. British Astron. Assoc., 106, 354

\bref Hoyle, F. 1946, MNRAS, 106, 343

\bref Kenney, J.D.P., van Gorkom, J.H., \& Vollmer, B. 2004, AJ, 127,
3361

\bref Kolmogorov, A. 1941, Dokl. Akad. Nauk SSSR, 30, 301

\bref Kuiper, 1935, Science, 82, 52

\bref Hartmann, J. 1904, ApJ, 19, 268

\bref Herbig, G.H. 1948, PhD Dissertation, Univ. California, Berkeley

\bref Herbig, George H. 1952, JRASC, 46, 222

\bref Herschel, W. 1811, Philosophical Transactions of the Royal
Society of London, 101, 269

\bref Holmes, A. 1927, The Age of the Earth, an Introduction to
Geological Ideas, London: Benn

\bref Joy, A.H. 1945, ApJ, 102, 168

\bref Hoyle, F. 1953, ApJ, 118, 513

\bref Hubble, E. 1929, Proc. Nat. Acad. of Sci. of the United States of
America, 15, 168

\bref Ivanenko, D., \& Pomeranchuk, I. 1944, Phys. Rev. 65, 343

\bref Jeans, J. 1929, Astronomy and Cosmogony, Cambridge Univ. Press,
p. 322

\bref Jeffreys, H. 1929, MNRAS, 89, 731

\bref Jenkins, E.B., \& Meloy, D.A. 1974, ApJ, 193, 121

\bref Lemaitre, G. Annales de la Soci\'et\'e Scientifique de Bruxelles,
47, 49

\bref Lilley, A. E. 1955, ApJ, 121, 559

\bref Lindblad, B. 1927, MNRAS, 87, 553

\bref Lyttleton, R.A. 1936, MNRAS, 96, 559

\bref McKee, C.F., \& Ostriker, J.P. 1977, ApJ, 218, 148

\bref Mestel, L., \& Spitzer, L., Jr. 1956, MNRAS, 116, 503

\bref Minkowski, R. 1955, in Gas Dynamics of Cosmic Clouds, ed. J.M.
Burgers \& H.C. van de Hulst, pp. 3--12. Amsterdam: North Holland

\bref Morgan, W. W., Whitford, A. E., \& Code, A. D. 1953, ApJ, 118,
318

\bref Muller, C. A., \& Oort, J. H. 1951, Nature, 168, 357

\bref M\"unch, G. 1953, PASP, 65, 179

\bref Oort, J.H. 1927, BAN, 3, 275

\bref Oort, J.H. 1932, BAN, 6, 249

\bref Oort, J.H. 1940, ApJ, 91, 273

\bref Oort, J. H. 1955, IAUS, 2, 147

\bref Oort, J.H., \& Spitzer, L., Jr. 1955, ApJ, 121, 6

\bref \"Opik, E. J., 1953, IrAJ, 2, 219

\bref Roman, N.G. 1950, ApJ, 112, 554

\bref  Routly, P.M., \& Spitzer, L., Jr. 1952, ApJ, 115, 227

\bref Russell, H.N. 1934 {\it The solar system and its origin}, New
York: Macmillan

\bref Salpeter, E.E. 1952, ApJ, 115, 326

\bref Salpeter, E.E. 1955, ApJ, 121, 161

\bref Schwarzschild, M., \& Schwarzschild, B. 1950, ApJ, 112, 248

\bref Schwarzschild, M., Spitzer, L., Jr., \& Wildt, R. 1951, ApJ, 114,
398

\bref Schwarzschild, M., \& Spitzer, L., Jr. 1953, Obs., 73, 77

\bref Seares, F.H. 1922, ApJ, 55, 165

\bref Smart, W.M. 1938, Stellar Dynamics, Cambridge: Cambridge Univ.

\bref Spitzer, L., Jr. 1936, MNRAS, 96, 794

\bref Spitzer, L., Jr. 1938a, PhD Thesis, Princeton Univ., American
Doctoral Dissertations, Source code: W1938, p. 5

\bref Spitzer, J., Jr. 1938b, ApJ, 87, 1

\bref Spitzer, L., Jr. 1939a, ApJ, 90, 494

\bref Spitzer, L., Jr. 1939b, ApJ, 90, 675

\bref Spitzer, L., Jr. 1940, ApJ, MNRAS, 100, 396

\bref Spitzer, L., Jr. 1941a, ApJ, 93, 369

\bref Spitzer, L., Jr. 1941b, ApJ, 94, 232

\bref Spitzer, L. Jr. 1942, ApJ, 95, 329

\bref Spitzer, L., Jr. 1950, ApJ, 111, 593

\bref Spitzer, L., Jr. 1952, ApJ, 116, 299

\bref Spitzer, L., Jr. 1954, ApJ, 120, 1

\bref Spitzer, L., Jr. 1956, ApJ, 124, 20

\bref Spitzer, L., Jr. 1958a, ApJ, 127, 17

\bref Spitzer, L., Jr. 1958b, Ricerche Astronomiche, Vol. 5, Specola
Vaticana, Proceedings of a Conference at Vatican Observatory, ed.
D.J.K. O'Connell, Amsterdam: North-Holland, p.445

\bref Spitzer, L., Jr. 1968 {\it Diffuse Matter in Space}, New York:
Interscience

\bref Spitzer, L., Jr. 1988 {\it Dynamical Evolution of Globular
Clusters}, Princeton University Press

\bref Spitzer, L., Jr., \& Baade, W. 1950, AJ, 55, 183

\bref Spitzer, L., Jr., \& Tukey, John W. 1951, ApJ, 114, 187

\bref Spitzer, L., Jr., \& Schwarzschild, M. 1951, ApJ, 114, 385

\bref Spitzer, L., Jr., \& Greenstein, J.L. 1951, ApJ, 114, 407

\bref Spitzer, L., Jr., \& Oke, J. B. 1952, ApJ, 115, 222

\bref Spitzer, L., Jr., \& Skumanich, A. 1952, ApJ, 116, 452

\bref Spitzer, L., Jr., \& Jenkins, E. B. 1975, ARA\&A, 13, 133

\bref Str\"omberg, G. 1925, ApJ, 61, 363

\bref Str\"omberg, G. 1934, ApJ, 79, 460

\bref Str\"omgren, B. 1939, ApJ, 89, 526

\bref Struve, O., \& Elvey, C. T. 1938, ApJ, 88, 364

\bref Teixeira, P.S., Lada, C.J., \& Alves, J.F. 2005, ApJ, 629, 276

\bref Trimble, V. 2009, ``Reviews in Modern Astronomy,'' volume 21 of
the Astronomische Gesellschaft, Wiley-VCH, ed. Siegfried Roeser

\bref Trumpler, R.J. 1934, PASP, 46, 208

\bref von Hoerner S. 1951, Z. Astrophys, 30, 17

\bref von Weizs\"acker, C.F. 1948, Naturwiss, 35, 188

\bref von Weizs\"acker, C.F. 1949, Problems of Cosmical Aerodynamics
(Dayton, Ohio: Central Air Documents Office); Proceedings of the Paris
Symposium on the Motion of Gaseous masses of Cosmical Dimensions,
chapter 22

\bref von Weizs\"acker C.F., 1951, ApJ, 114, 165

\bref Whipple, F. 1946, ApJ, 104, 1

\bref Wilson O.C., Minich G., Flather E., Coffeen M.F. 1959, ApJS, 4,
199

\bref Wilson, R.W., Jefferts, K.B., \& Penzias, A.A. 1970, ApJ, 161,
L43

\bref Zwicky, F. 1941, ApJ, 93, 411

\bref Zwicky, F. 1953, PASP, 65, 205

}

\vfill

\end{document}